\documentclass[reprint,aip, apl]{revtex4-1}
\usepackage{units, color, amssymb, sidecap}
\usepackage{graphicx}

\begin{document}


\title{Nanoimplantation and Purcell enhancement of single NV centers in photonic crystal cavities in diamond} 




\author{Janine Riedrich-M\"oller}
\affiliation{Universit\"at des Saarlandes, Fachrichtung 7.2 (Experimentalphysik), Campus E 2.6, 66123 Saarbr\"ucken, Germany}

\author{S\'ebastien Pezzagna} 
\affiliation{\mbox{Universit\"at Leipzig, Institut f\"ur Experimentalphysik II,  Linn\'estra{\ss}e 5, 04103 Leipzig, Germany}}

\author{Jan Meijer}
\affiliation{\mbox{Universit\"at Leipzig, Institut f\"ur Experimentalphysik II,  Linn\'estra{\ss}e 5, 04103 Leipzig, Germany}}

\author{Christoph Pauly}
\affiliation{Universit\"at des Saarlandes, Fachrichtung 8.4 (Materialwissenschaft und Werkstofftechnik), Campus D 3.3, 66123 Saarbr\"ucken, Germany}

\author{Frank M\"ucklich}
\affiliation{Universit\"at des Saarlandes, Fachrichtung 8.4 (Materialwissenschaft und Werkstofftechnik), Campus D 3.3, 66123 Saarbr\"ucken, Germany}

\author{Matthew Markham}
\affiliation{\mbox{Element Six Ltd., Global Innovation Centre, Fermi Avenue, Harwell Oxford, Didcot, OX11 0QR, UK}}

\author{Andrew M. Edmonds}
\affiliation{\mbox{Element Six Ltd., Global Innovation Centre, Fermi Avenue, Harwell Oxford, Didcot, OX11 0QR, UK}}

\author{Christoph Becher}
\email[]{christoph.becher@physik.uni-saarland.de}
\affiliation{Universit\"at des Saarlandes, Fachrichtung 7.2 (Experimentalphysik), Campus E 2.6, 66123 Saarbr\"ucken, Germany}


\date{\today}

\begin{abstract}
We present the controlled creation of single nitrogen-vacancy (NV) centers via ion implantation at the center of a photonic crystal cavity which is fabricated in an ultrapure, single crystal diamond membrane. High-resolution placement of NV centers is achieved using collimation of a \unit[5]{keV}-nitrogen ion beam through a pierced tip of an atomic force microscope (AFM). 
We demonstrate coupling of the implanted NV centers' broad band fluorescence to a cavity mode and observe Purcell enhancement of the spontaneous emission. The results are in good agreement with a master equation model for the cavity coupling.
\end{abstract}

\maketitle 

The nitrogen-vacancy (NV) center \cite{Doherty2013} in diamond has been successfully implemented as solid state quantum bit that meets all essential requirements for quantum information processing such as optical initialization, control and readout of the spin state.  The challenge remains to extend the quantum system from a small number of qubits to large scale networks. Seminal experiments already demonstrated remote entanglement between individual NV centers via two-photon quantum interference \cite{Sipahigil2012,Bernien2013}. The hitherto poor rate of  entanglement events \cite{Sipahigil2012,Bernien2013} could be strongly increased by coupling the NV centers to optical microcavities. The effects range from enhancement and spectral reshaping of the NV spectrum over cavity-enhanced spin state readout \cite{Young2009} to cavity mediated  entanglement between two NV centers \cite{Wolters2014}. Photonic crystal (PhC) cavities directly fabricated in diamond are ideal for color center-cavity coupling experiments as they exhibit high Q-factors and extremely small mode volumes. For solid state systems, it is however challenging to precisely place the emitter in the maximum of the cavity electric field to achieve optimum coupling.

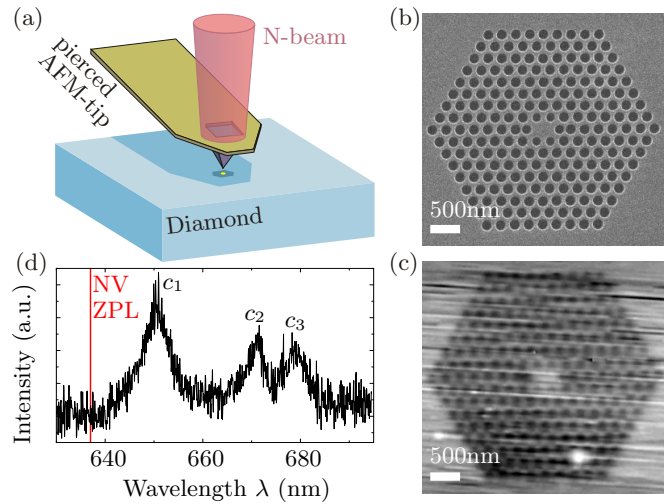
\begin{figure}
\centering
\begingroup%
  \makeatletter%
  \providecommand\color[2][]{%
    \errmessage{(Inkscape) Color is used for the text in Inkscape, but the package 'color.sty' is not loaded}%
    \renewcommand\color[2][]{}%
  }%
  \providecommand\transparent[1]{%
    \errmessage{(Inkscape) Transparency is used (non-zero) for the text in Inkscape, but the package 'transparent.sty' is not loaded}%
    \renewcommand\transparent[1]{}%
  }%
  \providecommand\rotatebox[2]{#2}%
  \ifx\svgwidth\undefined%
    \setlength{\unitlength}{245bp}%
    \ifx\svgscale\undefined%
      \relax%
    \else%
      \setlength{\unitlength}{\unitlength * \real{\svgscale}}%
    \fi%
  \else%
    \setlength{\unitlength}{\svgwidth}%
  \fi%
  \global\let\svgwidth\undefined%
  \global\let\svgscale\undefined%
  \makeatother%
  \begin{picture}(1,0.74264728)%
    \put(0,0){\includegraphics[width=\unitlength]{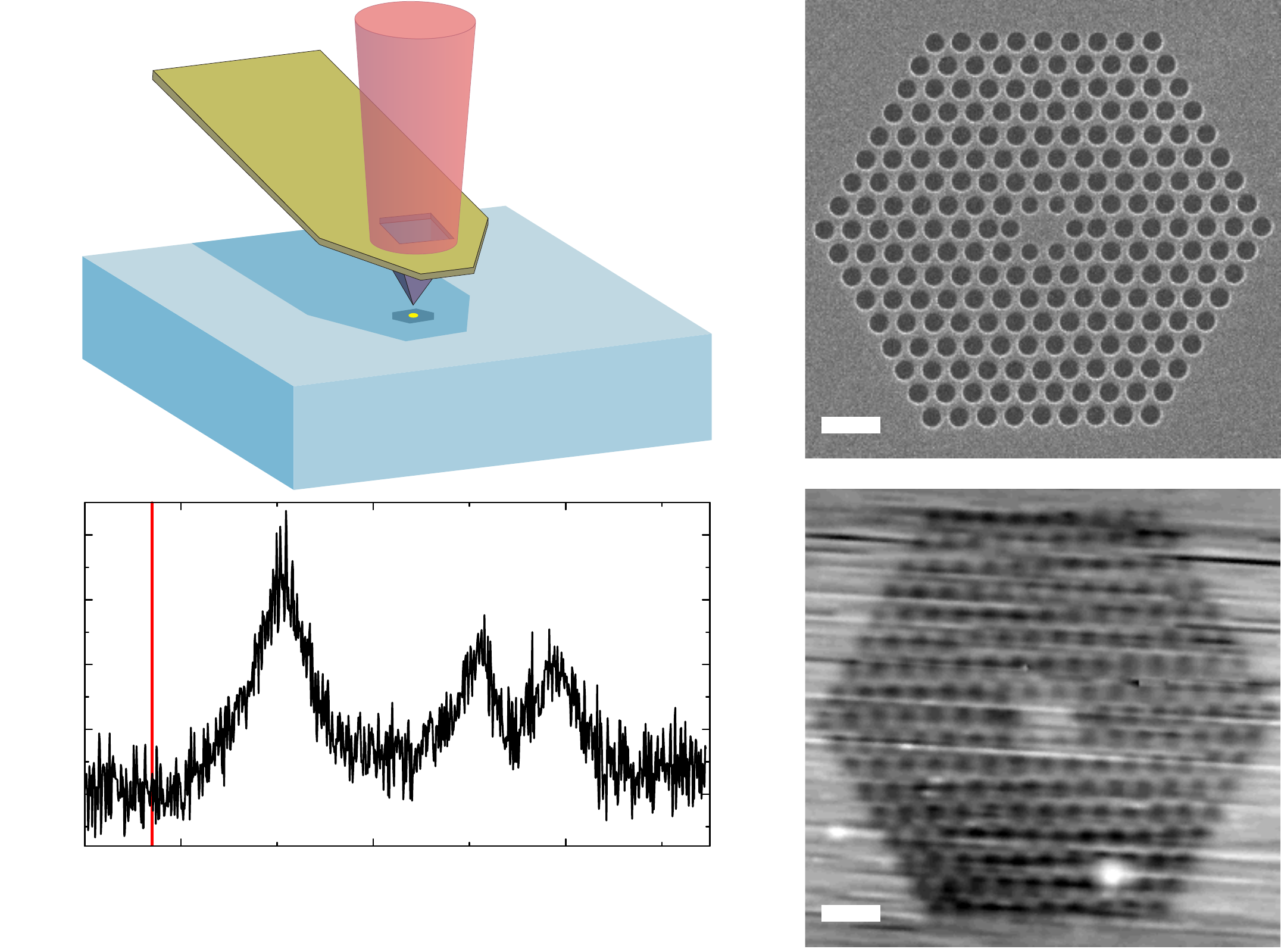}}%
    \put(0.38351654,0.69581111){\color[rgb]{0.72156863,0.40392157,0.4745098}\makebox(0,0)[lb]{\smash{N-beam}}}%
    \put(0.06259901,0.6953707){\color[rgb]{0.12156863,0.10196078,0.09019608}\rotatebox{-45.42252681}{\makebox(0,0)[lb]{\smash{pierced}}}}%
    \put(0.03719121,0.66978708){\color[rgb]{0.12156863,0.10196078,0.09019608}\rotatebox{-45.42252681}{\makebox(0,0)[lb]{\smash{AFM-tip}}}}%
    \put(0.23646254,0.40292265){\color[rgb]{0.12156863,0.10196078,0.09019608}\rotatebox{6.81638665}{\makebox(0,0)[lb]{\smash{Diamond}}}}%
    \put(-0.0011724,0.72437142){\color[rgb]{0.12156863,0.10196078,0.09019608}\makebox(0,0)[lb]{\smash{(a)}}}%
    \put(0.57777859,0.72437142){\color[rgb]{0.12156863,0.10196078,0.09019608}\makebox(0,0)[lb]{\smash{(b)}}}%
    \put(0.64082366,0.42951813){\color[rgb]{1,1,1}\makebox(0,0)[lb]{\smash{500nm}}}%
    \put(0.57777859,0.34838475){\color[rgb]{0.12156863,0.10196078,0.09019608}\makebox(0,0)[lb]{\smash{(c)}}}%
    \put(0.0031112,0.34838475){\color[rgb]{0.12156863,0.10196078,0.09019608}\makebox(0,0)[lb]{\smash{(d)}}}%
    \put(0.11796777,0.045538718){\makebox(0,0)[lb]{\smash{640}}}%
    \put(0.26814853,0.045538718){\makebox(0,0)[lb]{\smash{660}}}%
    \put(0.41832928,0.045538718){\makebox(0,0)[lb]{\smash{680}}}%
    \put(0.41748911,0.25942117){\makebox(0,0)[lb]{\smash{$c_3$}}}%
    \put(0.35531638,0.27146415){\makebox(0,0)[lb]{\smash{$c_2$}}}%
    \put(0.02586391,0.08002848){\rotatebox{90}{\makebox(0,0)[lb]{\smash{Intensity (a.u.)}}}}%
    \put(0.16827307,0.0000000){\makebox(0,0)[lb]{\smash{Wavelength $\lambda$ (nm)}}}%
    \put(0.22960563,0.31929037){\makebox(0,0)[lb]{\smash{$c_1$}}}%
    \put(0.64082366,0.05032743){\color[rgb]{1,1,1}\makebox(0,0)[lb]{\smash{500nm}}}%
		\put(0.12193561,0.31352118){\color[rgb]{1,0,0}\makebox(0,0)[lt]{\smash{NV}}}%
		\put(0.12193561,0.27352118){\color[rgb]{1,0,0}\makebox(0,0)[lt]{\smash{ZPL}}}%
  \end{picture}%
\endgroup%
\caption{Nanoimplantation process of nitrogen ions into diamond-based photonic crystal cavities:  (a) Schematic diagram of the nanoimplanter setup that combines collimation and positioning of a \unit[5]{keV} nitrogen ion beam with an AFM. A small hole in the AFM tip serves as an aperture for the ion beam. (b) SEM image and (c) AFM image of a fabricated M1-cavity. (d) M1-cavity spectrum prior implantation reveals three cavity modes $c_1$, $c_2$ and $c_3$ at 653, 670 and \unit[681]{nm} close to the theoretical NV ZPL at \unit[637]{nm} (red line). \label{fig1}}
\end{figure}
Past experiments that demonstrated coupling of single NV centers to PhC cavities \cite{Faraon2012, Hausmann2013} have largely relied on random positioning. 
Controlled lateral positioning and emitter-cavity coupling  has recently been achieved via a tailored fabrication process of a PhC around a single silicon-vacancy center in diamond \cite{Riedrich-Moller2014b}. Here, we pursuit the complementary approach based on targeted implantation of NV centers into pre-defined cavities in diamond. In recent years, several techniques for spatially selective formation of single NV centers in bulk diamond have been developed involving focused nitrogen ion beam  \cite{Lesik2013}, implantation through pierced AFM-tips \cite{Meijer2008,Pezzagna2010} and through small apertures in e-beam resist \cite{Toyli2010, Spinicelli2011}, mica foils \cite{Pezzagna2011a}, and silicon masks \cite{Bayn2015}. Using the silicon mask simultaneously as an etch mask would allow for controlled emitter-cavity placement \cite{Schroder2014a}. 

In our experiment, we achieve high resolution implantation of NV centers within two-dimensional diamond-based PhC cavities using a combined system of a nitrogen ion beam and an atomic force microscope (Fig. \ref{fig1}(a)) that allows for collimation and lateral positioning of the ion beam \cite{Meijer2008,Pezzagna2010}. 
We verify the successful formation  of a small number of  NV centers and demonstrate Purcell enhancement of the broad NV emission when coupled to the confined cavity field.
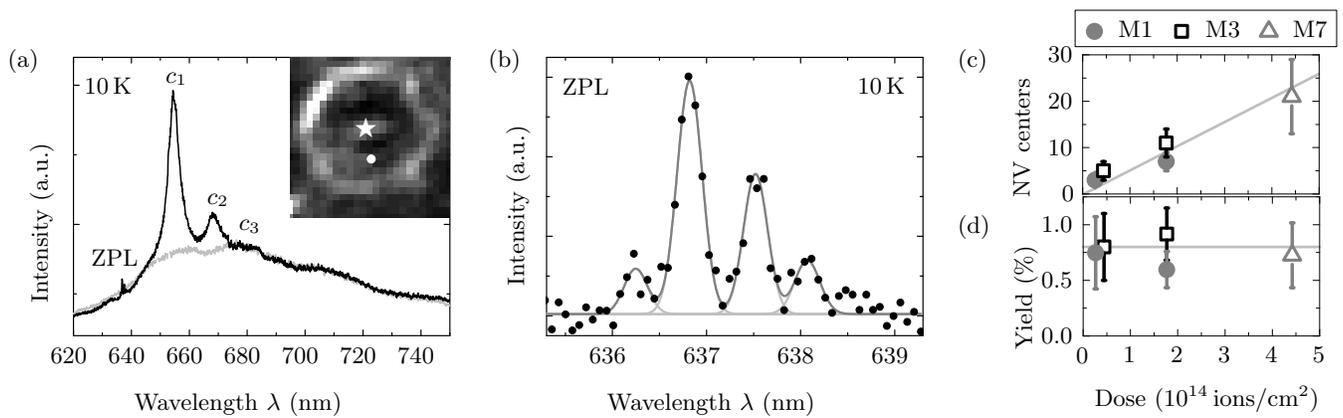
\begin{figure*}
\centering
\begingroup%
  \makeatletter%
  \providecommand\color[2][]{%
    \errmessage{(Inkscape) Color is used for the text in Inkscape, but the package 'color.sty' is not loaded}%
    \renewcommand\color[2][]{}%
  }%
  \providecommand\transparent[1]{%
    \errmessage{(Inkscape) Transparency is used (non-zero) for the text in Inkscape, but the package 'transparent.sty' is not loaded}%
    \renewcommand\transparent[1]{}%
  }%
  \providecommand\rotatebox[2]{#2}%
  \ifx\svgwidth\undefined%
    \setlength{\unitlength}{500bp}%
    \ifx\svgscale\undefined%
      \relax%
    \else%
      \setlength{\unitlength}{\unitlength * \real{\svgscale}}%
    \fi%
  \else%
    \setlength{\unitlength}{\svgwidth}%
  \fi%
  \global\let\svgwidth\undefined%
  \global\let\svgscale\undefined%
  \makeatother%
  \begin{picture}(1,0.30663775)%
    \put(0,0){\includegraphics[width=\unitlength]{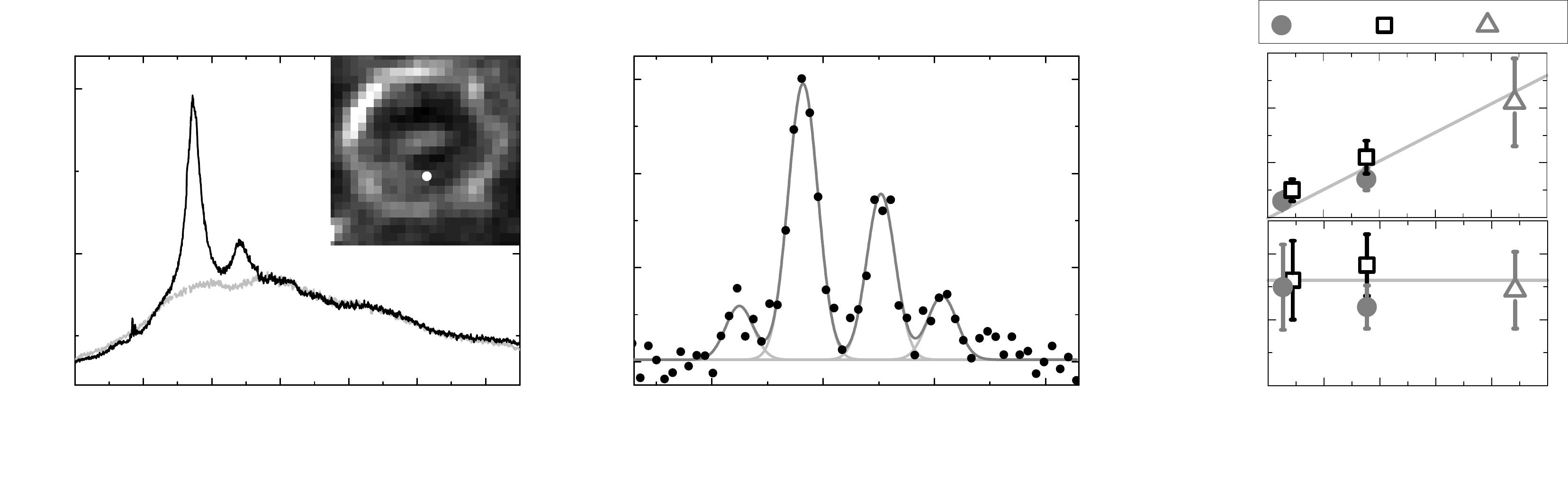}}%
    \put(0.43661678,0.04013226){\makebox(0,0)[lb]{\smash{636}}}%
    \put(0.50760559,0.04013226){\makebox(0,0)[lb]{\smash{637}}}%
    \put(0.57846854,0.04013226){\makebox(0,0)[lb]{\smash{638}}}%
    \put(0.64945735,0.04013226){\makebox(0,0)[lb]{\smash{639}}}%
    \put(0.03044499,0.04013226){\makebox(0,0)[lb]{\smash{620}}}%
    \put(0.07412073,0.04013226){\makebox(0,0)[lb]{\smash{640}}}%
    \put(0.11779647,0.04013226){\makebox(0,0)[lb]{\smash{660}}}%
    \put(0.16134634,0.04013226){\makebox(0,0)[lb]{\smash{680}}}%
    \put(0.20502208,0.04013226){\makebox(0,0)[lb]{\smash{700}}}%
    \put(0.24869782,0.04013226){\makebox(0,0)[lb]{\smash{720}}}%
    \put(0.29237356,0.04013226){\makebox(0,0)[lb]{\smash{740}}}%
    \put(0.79269364,0.1622331){\makebox(0,0)[lb]{\smash{0}}}%
    \put(0.7822467,0.19726307){\makebox(0,0)[lb]{\smash{10}}}%
    \put(0.7822467,0.23216702){\makebox(0,0)[lb]{\smash{20}}}%
    \put(0.7822467,0.26719698){\makebox(0,0)[lb]{\smash{30}}}%
    \put(0.80351817,0.04013226){\makebox(0,0)[lb]{\smash{0}}}%
    \put(0.83913844,0.0407623){\makebox(0,0)[lb]{\smash{1}}}%
    \put(0.87475871,0.0407623){\makebox(0,0)[lb]{\smash{2}}}%
    \put(0.91050485,0.0407623){\makebox(0,0)[lb]{\smash{3}}}%
    \put(0.94612513,0.0407623){\makebox(0,0)[lb]{\smash{4}}}%
    \put(0.9817454,0.0407623){\makebox(0,0)[lb]{\smash{5}}}%
    \put(0.7773379,0.0551271){\makebox(0,0)[lb]{\smash{0.0}}}%
    \put(0.7773379,0.09708745){\makebox(0,0)[lb]{\smash{0.5}}}%
    \put(0.7773379,0.13904781){\makebox(0,0)[lb]{\smash{1.0}}}%
    \put(0.71420936,0.13854378){\makebox(0,0)[lb]{\smash{(d)}}}%
    \put(0.63875868,0.24350766){\makebox(0,0)[lb]{\smash{\unit[10]{K}}}}%
    \put(0.38740291,0.09910357){\rotatebox{90}{\makebox(0,0)[lb]{\smash{Intensity (a.u.)}}}}%
    \put(0.45008452,0.00522831){\makebox(0,0)[lb]{\smash{Wavelength $\lambda$ (nm)}}}%
    \put(0.41622638,0.24275161){\makebox(0,0)[lb]{\smash{ZPL}}}%
    \put(0.14863381,0.16046901){\makebox(0,0)[lb]{\smash{$c_2$}}}%
    \put(0.17263381,0.14046901){\makebox(0,0)[lb]{\smash{$c_3$}}}%
    \put(0.02843112,0.08587283){\rotatebox{90}{\makebox(0,0)[lb]{\smash{Intensity (a.u.)}}}}%
    \put(0.0935042,0.00522831){\makebox(0,0)[lb]{\smash{Wavelength $\lambda$ (nm)}}}%
    \put(0.117293,0.25207614){\makebox(0,0)[lb]{\smash{$c_1$}}}%
    \put(0.06128233,0.11346837){\makebox(0,0)[lb]{\smash{ZPL}}}%
    \put(0.05612179,0.24388568){\makebox(0,0)[lb]{\smash{\unit[10]{K}}}}%
    \put(0.83536244,0.28622405){\makebox(0,0)[lb]{\smash{M1}}}%
    \put(0.90106485,0.28622405){\makebox(0,0)[lb]{\smash{M3}}}%
    \put(0.96676726,0.28622405){\makebox(0,0)[lb]{\smash{M7}}}%
    \put(0.36046744,0.26404681){\makebox(0,0)[lb]{\smash{(b)}}}%
    \put(0.76852723,0.16550929){\rotatebox{90}{\makebox(0,0)[lb]{\smash{NV centers}}}}%
    \put(-0.00177688,0.26404681){\makebox(0,0)[lb]{\smash{(a)}}}%
    \put(0.71420936,0.26404681){\makebox(0,0)[lb]{\smash{(c)}}}%
    \put(0.7701635,0.05084286){\rotatebox{90}{\makebox(0,0)[lb]{\smash{Yield (\%)}}}}%
    \put(0.8175331,0.00812647){\makebox(0,0)[lb]{\smash{Dose (\unit[10$^{14}$]{ions/cm$^2$)}}}}%
		\put(0.2598818,0.21260287){\color[rgb]{1,1,1}\makebox(0,0)[lb]{\smash{\small$\bigstar$}}}%
  \end{picture}%
\endgroup%
\caption{Low-temperature spectroscopy after nanoimplantation into PhC cavities, extracted number of NV centers and creation yield: (a) Inset: PL scan of the M1-cavity after \unit[5]{keV}-nitrogen ion implantation at the lowest dose of \unit[$3\times10^{13}$]{ions/cm$^2$} (area:  \unit[$5\times 5$]{$\mu$m$^2$}).  The spectrum taken at the center of the PhC (black line, position marked by white star in PL scan) clearly reveals signature of NV centers around \unit[637]{nm} and three cavity modes $c_1$, $c_2$ and $c_3$  at longer wavelengths, which are not present in the reference spectrum taken \unit[1]{$\mu$m} off the cavity center (gray line, position marked by white dot in PL scan). (b) Detailed, background corrected spectrum around \unit[637]{nm} displays $3\pm 1$ distinct NV ZPLs with  Gaussian lineshapes and  linewidths of \unit[250]{GHz}.  Dots: measured data, gray lines: Gaussian fits. (c) Number of created NV centers and their associated (d) production yield as a function of the applied dose upon $^{15}$N$^+$ ion implantation into M1, M3 and M7 PhC cavities. \label{fig2}}
\end{figure*}
The PhCs consist of a triangular lattice of air holes milled in a single crystal diamond membrane with a refractive index of $n=2.4$. The cavity is introduce by a one-, three- or seven-hole defect at the center, referred to as M1-, M3- or M7-cavity, respectively. 
For membrane preparation a high purity synthetic diamond (\unit[$<5$]{ppb} nitrogen concentration) was synthesized using microwave assisted chemical vapor deposition. The as grown (001) single crystal diamond was processed using standard diamond lapidary for bulk material removal followed by scaife polishing to thin the diamond to \unit[10]{$\mu$m}. The membrane is bonded on a silicon substrate via
a spin-on-glass adhesion layer (hydrogen silsesquioxane, Dow corning XR-1541). The silicon substrate has been partially removed in order to obtain a free-standing diamond membrane which is subsequently thinned to \unit[220]{nm} using reactive ion etching in an oxygen plasma and patterned with an array of air holes using focused ion beam milling with \unit[30]{keV}-Ga$^+$ ions \cite{Riedrich-Moller2012}. As a final step, the sample is annealed at 800$^\circ$C for \unit[2]{h} in vacuum and is thoroughly cleaned in a boiling mixture of nitric, sulfuric, and perchloric acid. 

The lattice constant $a = $ \unit[$220-240$]{nm} and air hole radii $R = $ \unit[$80-83$]{nm} of the M1-, M3- and M7-cavities are chosen such that the resonant modes are close to the design wavelength of \unit[637]{nm} with experimental quality factors of $Q=150-1200$ and mode volumes of $V \approx 1(\lambda/n)^3$. The hole radii of the M3- and M7-\-cavity are uniform in size, whereas the next neighbor holes around the M1-defect have been displaced and reduced in size to optimize the cavity $Q$ \cite{Kreuzer2008} (c.f. Figs. \ref{fig1}(b,c)).

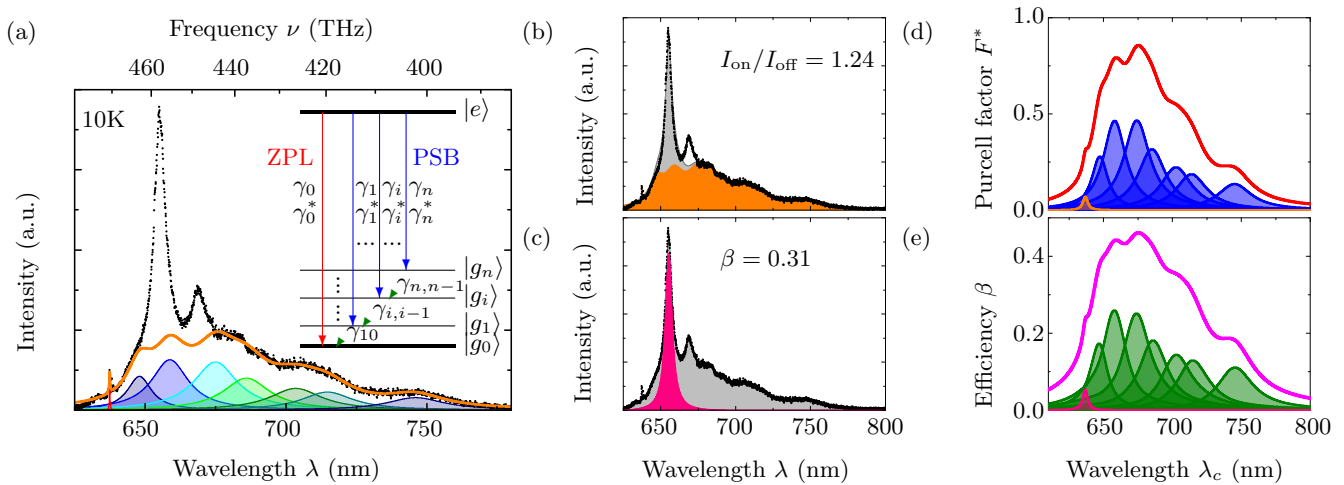
\begin{figure*}
\centering
\begingroup%
  \makeatletter%
  \providecommand\color[2][]{%
    \errmessage{(Inkscape) Color is used for the text in Inkscape, but the package 'color.sty' is not loaded}%
    \renewcommand\color[2][]{}%
  }%
  \providecommand\transparent[1]{%
    \errmessage{(Inkscape) Transparency is used (non-zero) for the text in Inkscape, but the package 'transparent.sty' is not loaded}%
    \renewcommand\transparent[1]{}%
  }%
  \providecommand\rotatebox[2]{#2}%
  \ifx\svgwidth\undefined%
    \setlength{\unitlength}{500bp}%
    \ifx\svgscale\undefined%
      \relax%
    \else%
      \setlength{\unitlength}{\unitlength * \real{\svgscale}}%
    \fi%
  \else%
    \setlength{\unitlength}{\svgwidth}%
  \fi%
  \global\let\svgwidth\undefined%
  \global\let\svgscale\undefined%
  \makeatother%
  \begin{picture}(1,0.36024425)%
    \put(0,0){\includegraphics[width=\unitlength]{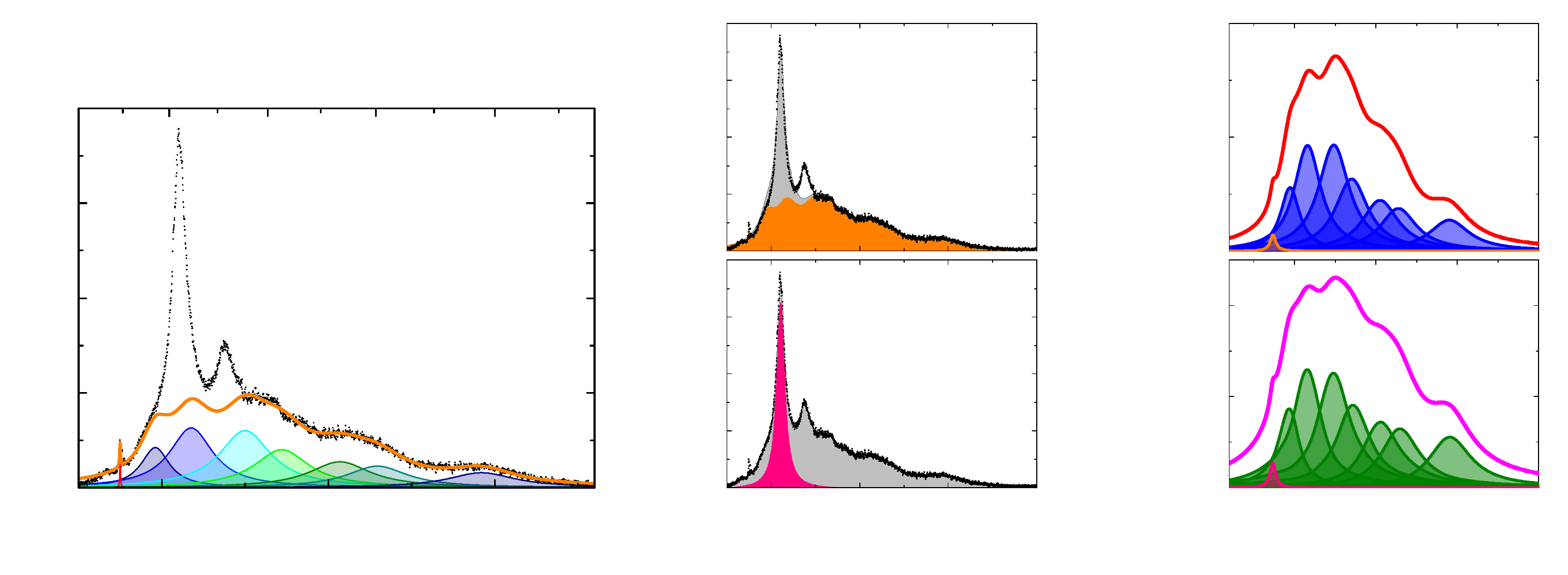}}%
    \put(0.81293658,0.03731239){\makebox(0,0)[lb]{\smash{650}}}%
    \put(0.86530168,0.03731239){\makebox(0,0)[lb]{\smash{700}}}%
    \put(0.91766679,0.03731239){\makebox(0,0)[lb]{\smash{750}}}%
    \put(0.97003189,0.03731239){\makebox(0,0)[lb]{\smash{800}}}%
    \put(0.75749978,0.04930593){\makebox(0,0)[lb]{\smash{0.0}}}%
    \put(0.75749978,0.10781097){\makebox(0,0)[lb]{\smash{0.2}}}%
    \put(0.75749978,0.16631602){\makebox(0,0)[lb]{\smash{0.4}}}%
    \put(0.75749978,0.20112652){\makebox(0,0)[lb]{\smash{0.0}}}%
    \put(0.75749978,0.27425783){\makebox(0,0)[lb]{\smash{0.5}}}%
    \put(0.75749978,0.34738913){\makebox(0,0)[lb]{\smash{1.0}}}%
    \put(0.08626104,0.03614229){\makebox(0,0)[lb]{\smash{650}}}%
    \put(0.19318532,0.03614229){\makebox(0,0)[lb]{\smash{700}}}%
    \put(0.3001096,0.03614229){\makebox(0,0)[lb]{\smash{750}}}%
    \put(0.67339381,0.18035723){\makebox(0,0)[lb]{\smash{(e)}}}%
    \put(0.67339381,0.3353956){\makebox(0,0)[lb]{\smash{(d)}}}%
    \put(0.38728726,0.18035723){\makebox(0,0)[lb]{\smash{(c)}}}%
    \put(0.38728726,0.3353956){\makebox(0,0)[lb]{\smash{(b)}}}%
    \put(0.05569035,0.26899237){\makebox(0,0)[lb]{\smash{10K}}}%
    \put(0.01956135,0.099474){\rotatebox{90}{\makebox(0,0)[lb]{\smash{Intensity (a.u.)}}}}%
    \put(0.12399902,0.00542714){\makebox(0,0)[lb]{\smash{Wavelength $\lambda$ (nm)}}}%
    \put(-0.00164798,0.3353956){\makebox(0,0)[lb]{\smash{(a)}}}%
    \put(0.09079545,0.30848328){\makebox(0,0)[lb]{\smash{460}}}%
    \put(0.15427717,0.30848328){\makebox(0,0)[lb]{\smash{440}}}%
    \put(0.22375601,0.30848328){\makebox(0,0)[lb]{\smash{420}}}%
    \put(0.3001096,0.30848328){\makebox(0,0)[lb]{\smash{400}}}%
    \put(0.12151241,0.3377358){\makebox(0,0)[lb]{\smash{Frequency $\nu$ (THz)}}}%
    \put(0.47695153,0.03731239){\makebox(0,0)[lb]{\smash{650}}}%
    \put(0.53385104,0.03731239){\makebox(0,0)[lb]{\smash{700}}}%
    \put(0.59075056,0.03731239){\makebox(0,0)[lb]{\smash{750}}}%
    \put(0.64765007,0.03731239){\makebox(0,0)[lb]{\smash{800}}}%
    \put(0.47768289,0.00484209){\makebox(0,0)[lb]{\smash{Wavelength $\lambda$ (nm)}}}%
    \put(0.43921355,0.06086067){\rotatebox{90}{\makebox(0,0)[lb]{\smash{Intensity (a.u.)}}}}%
    \put(0.53677647,0.16412208){\makebox(0,0)[lb]{\smash{$\beta = 0.31$}}}%
    \put(0.44067626,0.20946349){\rotatebox{90}{\makebox(0,0)[lb]{\smash{Intensity (a.u.)}}}}%
    \put(0.53677647,0.31448005){\makebox(0,0)[lb]{\smash{$I_{\text{on}}/I_{\text{off}} = 1.24$}}}%
    \put(0.74155621,0.05735037){\rotatebox{90}{\makebox(0,0)[lb]{\smash{Efficiency $\beta$}}}}%
    \put(0.79713928,0.00484209){\makebox(0,0)[lb]{\smash{Wavelength $\lambda_c$ (nm)}}}%
    \put(0.74155621,0.20624571){\rotatebox{90}{\makebox(0,0)[lb]{\smash{Purcell factor $F^*$}}}}%
		\put(0.22,0.104){\linethickness{1.5pt}\line(1,0){0.117}}
		\put(0.22,0.119){\line(1,0){0.117}}
		\put(0.22,0.140){\line(1,0){0.117}}
		\put(0.22,0.161){\line(1,0){0.117}}
		\put(0.22,0.28){\linethickness{1.5pt}\line(1,0){0.117}} 
		\put(0.343,0.101){$|g_0\rangle$} 
		\put(0.343,0.116){$|g_1\rangle$} 
		\put(0.343,0.137){$|g_i\rangle$} 
		\put(0.343,0.158){$|g_n\rangle$} 
		\put(0.343,0.277){$|e\rangle$} 
		\put(0.237,0.28){\color[rgb]{1,0,0}\vector(0,-1){0.176}}  
		\put(0.30,0.28){\color[rgb]{0,0,1}\vector(0,-1){0.119}}  
		\put(0.28,0.28){\color[rgb]{0,0,1}\vector(0,-1){0.14}}  
		\put(0.26,0.28){\color[rgb]{0,0,1}\vector(0,-1){0.161}}  
		\put(0.26,0.122){\color[rgb]{0,0.5,0}\vector(-2,-3){0.013}}  
		\put(0.28,0.137){\color[rgb]{0,0.5,0}\vector(-2,-3){0.013}}  
		\put(0.30,0.158){\color[rgb]{0,0.5,0}\vector(-2,-3){0.013}}  
		\put(0.255,0.1105){$\gamma_{10}$} 
		\put(0.275,0.127){$\gamma_{i,i-1}$} 
		\put(0.293,0.147){$\gamma_{n,n-1}$} 
		\put(0.195,0.24){\color[rgb]{1,0,0}{ZPL}} 
		\put(0.305,0.24){\color[rgb]{0,0,1}{PSB}} 
		\put(0.215,0.22){$\gamma_{0}$} 
		\put(0.215,0.20){$\gamma_{0}^{*}$} 
		\put(0.262,0.22){$\gamma_{1}$} 
		\put(0.262,0.20){$\gamma_{1}^{*}$} 
		\put(0.282,0.22){$\gamma_{i}$} 
		\put(0.282,0.20){$\gamma_{i}^{*}$} 
		\put(0.302,0.22){$\gamma_{n}$} 
		\put(0.302,0.20){$\gamma_{n}^{*}$} 
		\put(0.248,0.121){\rotatebox{90}{$...$}} 
		\put(0.248,0.142){\rotatebox{90}{$...$}} 
		\put(0.262,0.18){{$...$}} 
		\put(0.282,0.18){{$...$}} 
		\end{picture}%
		\endgroup%
\caption{Spectrally resolved Purcell enhancement of the NV emission via coupling to the M1-cavity mode: (a) From the measured NV/M1-cavity spectrum (black dots),  the bare, uncoupled NV emission (orange line) is estimated.  The bare spectrum is fitted with eight Lorentzians according to the (inset) multi-level model of the NV center including one excited state $|{e}\rangle$ and eight vibrational  ground states $|{g_i}\rangle, i\in[{0,7}]$. The parameters are the transition rates $\gamma_i$, the pure dephasing rate $\gamma^*$ and the relaxation rates $\gamma_{i,i-1}$ between vibrational ground state sublevels. (b) Experimental enhancement of the on resonance NV intensity $I_{\text{on}}$ (orange + gray) coupled to  cavity mode $c_1$ (disregarding modes $c_2$, $c_3$) compared  to the uncoupled case $I_{\text{off}}$ (orange).  (c) Experimental emission efficiency $\beta $ defined as the ratio of the intensity $I_{\text{on}}^{c_1}$ (pink) channeled into cavity mode $c_1$ to the total emission $I_{\text{on}}^{\text{tot}}$ (pink + gray).
Calculated (d) generalized Purcell factor $F^*$ and (e)  emission efficiency $\beta$ as a function of cavity
wavelength $\lambda_c$. The individual contributions of the ZPL and PSB to emitter-cavity coupling are shown by the filled curves.
 \label{fig3}}
\end{figure*}

The fabricated structures are investigated using a home-build confocal microscopy setup with a continuous-wave \unit[532]{nm} excitation laser where the sample is mounted in a continuous flow liquid-helium cryostat.
Figure \ref{fig1}(d) shows the room temperature photoluminescence (PL) spectrum of the fabricated M1-cavity revealing three  pronounced cavity modes $c_1$, $c_2$ and $c_3$ at 653, 670 and \unit[681]{nm}, respectively, but no signature of NV emission in the ultrapure diamond material.

For deterministic creation of NV centers within the PhC cavities, we first use an AFM to image the PhC structures (c.f. Figs. \ref{fig1}(c)).
A small hole (diameter \unit[$< 30$]{nm}) drilled in the AFM tip serves as an aperture 
for the ion beam that allows for high resolution
implantation \cite{Meijer2008,Pezzagna2010} of $^{15}$N$^+$ ions  with an energy of \unit[5]{keV} at the cavities' center at different doses of \unit[$0.3-4.4\times 10^{14}$]{ions/cm$^2$}. The low ion energy  is chosen to achieve a high spatial resolution of \unit[$< 15$]{nm} \cite{Pezzagna2010}.  According to Monte Carlo simulations (SRIM \cite{Srim2008}), the average implantation depth of the \unit[5]{keV}-nitrogen ions is \unit[8]{nm} with a small ion straggle of \unit[3]{nm}. 
After implantation, the diamond sample is annealed at $800^\circ$C for \unit[2]{h} in vacuum such that lattice vacancies diffuse in the diamond host material towards the implanted nitrogen ions to form optically active NV centers. Finally, the sample is cleaned again in a boiling acid mixture for \unit[8]{h} in order to oxidize any graphite-like residuals and to convert the NV centers to the negative charge state. 

Ensemble NV emission spectra taken at different reference spots implanted at high dose (\unit[$5\times 10^{14}$]{ions/cm$^2$}) aside the photonic structures reveal that up to 70\% of the NV centers are converted to their negative charge state after all oxidation steps. In our analysis, we take into account the PL intensities integrated in a spectral range of \unit[20]{nm} around the NV$^0$ ($\lambda =$ \unit[575]{nm}) and NV$^-$ zero-phonon line (ZPL) ($\lambda =$ \unit[637]{nm}) as well as varying detection efficiencies of our spectrometer, different absorption \cite{Beha2012a} and quantum \cite{GattoMonticone2013} efficiencies  of the two charge states.
In the following, we refer to the most abundant NV$^-$ center simply as NV center.

We verify the successful formation of NV centers within  PhC cavities using  confocal spectroscopy at \unit[10]{K}.  
Figure \ref{fig2} shows the PL scan and spectra of the M1-cavity after nitrogen ion implantation at the lowest dose of \unit[$3\times 10^{13}$]{ions/cm$^2$}. Besides the three cavity modes $c_1$, $c_2$ and $c_3$, a clear signature of NV ZPLs around \unit[637]{nm} is visible in the spectrum collected at the M1-cavity center (black spectrum in Fig. \ref{fig2}(a)). A zoom into the spectral region around \unit[637]{nm} (Fig. \ref{fig2}(b)) reveals $3\pm1$ narrow Gaussian-shaped lines with linewidths of $\Delta \nu_0 \approx$ \unit[250]{GHz}. At low temperature the linewidth is limited by spectral diffusion \cite{Wolters2013a}. Each line corresponds to the ZPL of a single NV center that has been created upon ion implantation and subsequent annealing. The implantation is solely restricted to the cavity center. Reference spectra collected \unit[1]{$\mu$m} off the cavity center (gray spectrum in Fig. \ref{fig2}(a)) do not show any signature of NV ZPLs, which verifies the high spatial resolution of the implantation process. 

We determine the number of NV centers created within each M1-, M3- and M7-cavity  that were implanted at various ion doses by  integrating the background corrected PL signal in the spectral range of \unit[$637\pm8$]{nm} and normalize it to the average ZPL intensity of a single NV center. As displayed in Fig. \ref{fig2}(c), the number of produced NV centers monotonically increases as a function of the applied implantation dose.  By dividing the number of NV centers by the amount of implanted nitrogen ions, we obtain the NV creation yield shown in Fig. \ref{fig2}(d). For the ion energy of \unit[5]{keV}, we find a creation yield of $0.8\pm 0.2\%$ that is constant over the whole range of implantation dose. 
The small creation yield is within the range of experimental observations of yields $< 0.1 \%$ \cite{Pezzagna2010a} to $\sim 25\%$ \cite{Antonov2014} which strongly depend on annealing and surface conditions. The creation yield is limited by loss of vacancies to the surface upon shallow implantation and surface effects possibly shifting the NV center charge state to NV$^+$.
From Fig. \ref{fig2}(c), we deduce an optimal dose of \unit[$1\times 10^{13}$]{ions/cm$^2$} at an ion energy of \unit[5]{keV} for deterministic creation of one single optically active NV center. 

In the following, we analyze the intensity enhancement of the NV emission at the resonant wavelength of the M1-cavity mode $c_1$. Thereby, we take into account that the mode $c_1$ preferentially overlaps with the  NV phonon side band (PSB) and not with the ZPL. As the resonance wavelengths and linewidths are known from the M1-cavity spectrum (Fig. \ref{fig1}) prior implantation, we can estimate the bare NV emission (orange line in Fig. \ref{fig3}(a)) without cavity modes. By comparing the integrated intensity of the cavity-enhanced emission $I_{\text{on}}$ (orange + gray area in Fig. \ref{fig3}(b)) to the bare spectrum $I_{\text{off}}$ (orange area), we find an experimental enhancement factor of $I_{\text{on}}/I_{\text{off}} = 1.24$. Here, we solely focus on the intensity increase by the dominant $c_1$  mode and disregard other modes $c_2$ and $c_3$. In addition the emission efficiency into the cavity mode is $\beta = I^{c_1}_{\text{on}}/I^{\text{tot}}_{\text{on}} = 0.31$, where $I^{c_1}_{\text{on}}$ is the emission channeled into mode $c_1$ (pink area in Fig. \ref{fig3}(c)) and $I^{\text{tot}}_{\text{on}}$ is the overall  intensity (pink + gray area).

To relate the resonant intensity enhancement of the NV PSB to a generalized Purcell factor $F^*$, we adopt the master equation model \cite{Albrecht2013, Auffeves2009} for broad-band emitter-cavity coupling. In analogy to \cite{Albrecht2013}, the NV emission is modeled as a multi-level system  (inset in Fig. \ref{fig3}(a)). 
The model input parameters 
are obtained from Lorentzian fits to the uncoupled NV spectrum (Fig.  \ref{fig3}(a)).  Solving the master equation model, we compute the generalized Purcell factor $F^*$ \cite{Albrecht2013, Auffeves2009} (Fig. \ref{fig3}(d)) and the associated emission factor $\beta = F^*/(1+F^*)$ (Fig. \ref{fig3}(e)) into the cavity when the resonant mode $c_1$ with $Q_{c_1}=160$ and $V_{c_1} = 1.1\, (\lambda/n)^3$ is tuned across the modeled NV spectrum. The individual contributions of the  ZPL and PSBs to the total emitter-cavity coupling are shown by the filled curves in Figs. \ref{fig3}(d,e). For simplicity, our analysis  assumes  unity quantum efficiency and perfect spatial and orientational overlap of the two NV dipoles with the cavity field. At the resonant wavelength $\lambda_{c_1} = $ \unit[653]{nm} of the M1-cavity mode (c.f. Fig. \ref{fig3}(d)), we find a theoretical Purcell enhancement of $1+F^* = 1.7$ and an emission efficiency $\beta = 0.42$ that result in a theoretical intensity increase $I_{\text{on}}/I_{\text{off}} = ((1+F^*)\,e_{1,2}\, \gamma+(1-e_{1,2})\gamma) / \gamma = 1.2$, considering that the $c_1$ mode preferentially overlaps with the first and second NV PSB with a relative contribution to the total NV emission of $e_{1,2} = 0.29$. These theoretical values are in excellent agreement with our experiment. 

In conclusion, we have demonstrated high resolution creation of a small number of NV centers at the center of diamond-based PhC cavities using collimated nanoimplantation of nitrogen ions through a pierced AFM-tip. For an ion energy of \unit[5]{keV}, we found a constant NV creation yield of 0.8\%, independent of implantation dose and cavity size. 
The lowest ion dose of \unit[$3\times 10^{13}$]{ions/cm$^2$} yielded $3\pm1 $ NV centers placed at the center of a M1-cavity. 
The emitter-cavity coupling leads to an intensity enhancement of $I_{\text{on}} / I_{\text{off}} = 1.24$ when the cavity mode is in resonance with the NV PSB and an emission efficiency into the cavity mode of $\beta = 0.31$ which is  
in very good agreement with theoretical predictions. 
From our experiment, we deduce an optimal dose of \unit[$1\times 10^{13}$]{ions/cm$^2$} for the targeted  creation of one single NV center within a PhC cavity.
The here demonstrated high resolution implantation of single NV centers within PhC cavities is an essential step towards scalable solid-state quantum networks \cite{Nemoto2013} or quantum repeaters \cite{Childress2005} based on NV nanocavity systems.

We cordially thank R. Albrecht and A. Bommer for helpful discussions, S. Wolff and C. Dautermann (Nano Structuring Center, University of Kaiserslautern) for assistance with the bonding process and A. Baur and M. Wandt (IMTEK, University of Freiburg) for deep reactive ion etching. This research has been partially funded by the European Community's Seventh Framework Programme (FP7/2007-2013) under Grant Agreement No. 618078 (WASPS) and No. 611143 (DIADEMS). EU funding for the project AME-Lab (European Regional Development Fund C/4-EFRE 13/2009/Br) for the FIB/SEM is  acknowledged. MM and AME additionally acknowledge finical support from the DARPA SPARQC.


%

\end{document}